\documentclass[a4paper,twocolumn,11pt,unpublished]{quantumarticle}
\makeatletter
\providecommand{\@afterenddocumenthook}{}
\makeatother
\pdfoutput=1
\usepackage[utf8]{inputenc}
\usepackage[english]{babel}
\usepackage[T1]{fontenc}
\usepackage{amsmath}
\usepackage{amssymb}
\usepackage{hyperref}
\usepackage{tikz}

\usepackage{lipsum}
\usepackage{geometry}
\usepackage{setspace}
\usepackage{titlesec}
\usepackage{quantikz}
\usepackage{float}
\usepackage{subcaption}
\usepackage{listings}
\usepackage{xcolor}
\usepackage{natbib}
\usepackage{orcidlink}

\begin{document}

\title{Gradient descent reliably finds depth- and gate-optimal circuits for generic unitaries}

\author{Alex Meiburg\orcidlink{0000-0002-4506-9146}}
\affiliation{Perimeter Institute for Theoretical Physics}
\affiliation{Institute for Quantum Computing, University of Waterloo}
\email{gdqcs@ohaithe.re}

\author{Janani Gomathi\orcidlink{0009-0004-5654-9382}}
\affiliation{Perimeter Institute for Theoretical Physics}
\affiliation{Institute for Information Processing (tnt/L3S), Leibniz Universität Hannover, Germany}
\email{jananirajagopal47@gmail.com}

\maketitle

\begin{abstract}
When the gate set has continuous parameters, synthesizing a unitary operator as a quantum circuit is, in principle, always possible using exact methods. However, efficiently finding depth- and gate-minimal circuits remains a major challenge. The landscape is very different for {\em compiled} unitaries, which arise from programming and typically have short circuits, as compared with {\em generic} unitaries, which use all parameters and typically require circuits of maximal size. Previous approaches based on random combinatorial search indicate a low success rate even when the circuit ansatz is nominally adequately parameterized, motivating the use of heavily overparameterized circuits. In this work, we present a gradient-based optimization framework that enables the synthesis of depth- and gate-optimal circuits for generic unitaries without overparameterization, even under restricted hardware connectivity. We prescribe parameter-optimal circuit skeletons and eliminate the need for random combinatorial search. We further show that the poor performance of earlier random-search approaches can be attributed to the inadvertent selection of parameter-deficient circuit topologies. By systematically avoiding such skeletons, our approach achieves reliable convergence while maintaining parameter efficiency.
\end{abstract}

\section{Introduction}
\par 
Unitary synthesis refers to the task of decomposing a unitary operator, represented as a matrix, into a series of quantum gates natively supported by hardware in a quantum computer - typically, one- and two-qubit gates. This can be done deterministically using algorithms like Quantum Shannon Decomposition \cite{Shende2006-qs}, but this will generally produce circuits that are far from optimal in size. Moreover, in general, finding an optimal decomposition is believed to be very hard \cite{Chia2021-pk,Hitchcock2015-ac}. The correct notion of optimality is also context dependent, for instance, there are works which focus on synthesizing Clifford circuits with a low CNOT count, synthesizing Clifford+T circuits with a low T count \cite{mosca20,Rietsch2024-mu}, or on accommodating limited hardware connectivity\cite{Li2019-dj,Zou2024-gg}.

In the setting where our available gate set includes real parameters (the gate set of arbitrary single-qubit unitaries + CNOT, for example), one approach is to posit a circuit "skeleton" of parameterized gates and then search over the real parameter space to minimize the error between the target and synthesized unitary. The resulting optimization problem is highly non-convex, but the earlier works \cite{Ashhab2022-hi,Ashhab2024-yw}, have shown some forms of success with gradient descent-based approaches, sometimes referred to as GRAPE \cite{Grape2, Khaneja2005-fy}. In the Quantum Machine Learning (QML) literature, works such as \cite{kiani2020learningunitariesgradientdescent} have also numerically investigated the question of whether a sufficiently parameterized landscape guarantees convergence of gradient-descent.

For unitaries that arise from logical programs, it is typical to expect a short circuit due to the presence of a particular structure or symmetry. But for a \textit{generic} unitary on $n$-qubits, a circuit of depth $\Theta(4^n/n)$ is required, as given by parameter counting. Even though we know that an optimal solution exists, we see that works such as \cite{Ashhab2024-yw} resorted to random combinatorial search and required overparameterization, especially as the system size increases, to guarantee a solution. Further, it suggested that gradient descent could do this sometimes at adequate parametrization, but only with some probability that rapidly fell to zero as the number of qubits grew. We find that the key to reaching the guaranteed optimal solution, in the parameter-optimal regime, is through one simple prescription for selecting the circuit skeleton, in which case gradient descent is able to find a depth-optimal circuit every time. These circuits naturally turn out to be optimal in terms of the number of entangling gates, and in the number of total gates, as well.

Our investigations are empirical, but strongly suggest that random combinatorial search, when adequately parametrized, fails only when certain bad circuit skeletons are selected, resulting in an \textit{effectively underparameterized} circuit. This is validated by further experiments on circuit topologies with connectivity constraints: gradient descent continued to find optimal-size circuits even in 1D or star-topology qubit layouts and never failed even across hundreds of trials. 

\section{Background}
Usually, there are two kinds of optimizations that one does in the case of transforming unitaries into circuits. When the unitary is already given as a circuit, the goal is to find a shorter or hardware-compatible circuit; this process is typically referred to as "compiling". Meanwhile, "synthesis" typically refers to when the unitary is given as the full $2^n \times 2^n$ matrix \cite{Rietsch2024-mu,Ashhab2024-yw,Ashhab2022-hi}. A common algorithm that performs exact synthesis for 1-qubit systems is a version of the Kliuchnikov–Maslov–Mosca (KMM) algorithm \cite{kliuchnikov2013fastefficientexactsynthesis} that uses algebraic number theory to break down a unitary into Clifford + T gates. 

Moreover, there are many different existing approaches to unitary synthesis for larger systems. One of the recent methods in approximate synthesis uses the process of reinforcement learning (RL), and in this context, unitary synthesis is framed as a sequential decision-making problem in which the RL agent constructs a quantum circuit one gate at a time. While it is promising, a major disadvantage of this approach is that the training is often unitary-specific. Although some recent works propose robust or generalized training schemes \cite{Rietsch2024-mu}, the approach remains difficult to scale to larger systems. Another decision-making approach used in unitary synthesis is Mixed-Integer Programming \cite{nagarajan2021quantumcircuitoptopensourceframeworkprovably}.  Here, hardware constraints, such as gate types and connectivity, are encoded as part of this problem. The objective is to determine an optimal circuit structure that implements the target unitary while respecting these constraints. Along with scalability issues, this method has the additional limitation that MIP problems are NP-hard problems at worst, which means that they may not be tractable in all cases.

Quantum compilers, such as the one provided by Qiskit, are responsible for transforming the entire high-level quantum algorithms into hardware executable circuits while satisfying the constraints of the target quantum device. This uses Quantum Shannon Decomposition \cite{Shende2006-qs}, which gives a straightforward way to synthesize an arbitrary unitary matrix into {\em a} circuit, but the resulting circuit can be far from optimal. In particular, the use of recursive decomposition means that it is not parameter-optimal, as there is no attempt to cancel out redundant parameters at each step. Various works have investigated algorithms to find shorter circuits, although the general problem is hard\cite{Chia2021-pk,Hitchcock2015-ac}. Special cases include synthesizing Clifford circuits with a low CNOT count, synthesizing Clifford+T circuits with a low T count\cite{Rietsch2024-mu}, or focusing on accommodating limited hardware connectivity\cite{Li2019-dj,Zou2024-gg}.
\par
In our approach, we aim to prescribe a parameter-optimal method for unitary synthesis using gradient descent. We consider the unitary synthesis problem as, given an $n$-qubit unitary matrix $U \in \mathbb{C}^{2^n \times 2^n}$, to produce a decomposition into a prescribed gate set. Up to an irrelevant global phase, $U$ has $4^n - 1$ real parameters, so if each individual gate has at most $p$ real parameters, a minimum of $\lceil{\frac{4^n - 1}{p}\rceil}$ gates will be needed for a generic unitary. A majority of effort has focused on identifying structured unitaries that can do much better than this bound, but generic unitaries still play an important role for holding input parameters in quantum physics or chemistry simulations, e.g. \cite{chawdhry24,Arrazola2022-oy}. Accordingly, we focus on the generic case, where the main question is whether the $\lceil{\frac{4^n - 1}{p}\rceil}$ bound can be tightly achieved. Existing numerical evidence \cite{kiani2020learningunitariesgradientdescent, Kandala2017HardwareEfficientVQE} has suggested that there should be no local minima preventing a perfect fit in the adequately parameterized regime, but in practice symmetries can mean that failure is quite possible if the circuit topology is not chosen carefully. We give a prescription to avoid this.

\section{Problem Setup and Method}

\subsection{Parametrization}
We aim to decompose any target unitary into a quantum circuit made of one-qubit and two-qubit gates, from a continuous gate set. We focus especially on the case where the CNOT gate is the unique two-qubit entangling gate, and arbitrary single-qubit unitaries are allowed, with three degrees of freedom each. Such a decomposition is called a hardware-efficient ansatz \cite{Kandala2017HardwareEfficientVQE} and can be expressed as \cite{Ashhab2022-hi},
\begin{align}
    U_{\operatorname{circ}} = S_\ell T_{\ell-1} \dots S_2 T_1 S_1
\end{align}
Here each $S_i = U_1 \otimes U_2..\otimes U_j..$, i.e., the tensor product of some single qubit unitaries, and the $T_i$ is some set of simultaneous CNOT gates. We call each $S_iT_i$ a layer, the final layer being just $S_\ell$, and we say this is an $\ell$-layer decomposition of the unitary. Each single-qubit $U_j$ is in turn parameterized by the three Euler angles, 
\begin{align}
    U_j(\theta_1,\theta_2,\theta_3) = R_z(\theta_2)R_y(\theta_1)R_z(\theta_3) 
\end{align}
where $R_y(\theta) = \text{exp}(-i\theta \text{Y}/2$) and $R_z(\theta) = \text{exp}(-i\theta \text{Z}/2)$. After fixing $T_i$, an appropriate circuit decomposition is sought by optimizing the parameters of these single-qubit unitaries so that the circuit implements the target unitary up to some precision. For useful quantum computing, an error (one minus fidelity $F$) in the range of $10^{-7} \sim 10^{-9}$ is a typical threshold; in our experiments, this is roughly the threshold of numerical error, and we assume that this indicates that there is indeed a perfect synthesis. Running with high precision numerics in a few cases seems to affirm this belief.

Each $T_i$ is made up of an arrangement of CNOTs, for example $T_i= \text{CNOT}_{1\rightarrow3}\otimes \text{CNOT}_{2 \rightarrow 4} \ldots$. Note that directionality is ultimately irrelevant, since $\text{CNOT}_{1\rightarrow 2}$ and $\text{CNOT}_{2\rightarrow 1}$ are equivalent under appropriate single-qubit transformations on either side. The pattern of CNOTs dictates what we call the "circuit skeleton", and as a discrete variable, it is held constant during gradient descent. Other works have considered searching over possible topologies, either by random search, brute-force enumeration, or relaxing the constraint to some continuous variable approximate constraint. In a sense, our main innovation is noticing that, for generic unitaries, this random search is unnecessary, and we can simply pick an ideal skeleton from the get-go using a simple design principle. We also give a rigorous proof that, as the number of qubits increases, the probability of a random topology succeeding becomes doubly exponentially small (see Appendix \ref{app:failure-prob}).

\subsection{Cost Function}
In order to perform gradient descent, we first define an appropriate cost function. Let the target unitary be $U_{\text{goal}}$ and the unitary implemented by the quantum circuit be denoted by $U_{\text{circ}}$. For optimal parameters, we must find that
\begin{equation}
    \text{D} =U_{\operatorname{goal}}U^{\dagger}_{\operatorname{circ}} \approx \mathbb{I}
    \label{deviation}
\end{equation}
up to a global phase.\\
Hence, the cost function must measure the deviation from $\mathbb{I}$ up to a global phase, which is the distance between the two unitaries. For ease of computation, we choose a trace-based distance measure. Therefore,
\begin{equation}
    C = N - \text{abs}(\text{Tr(D))}
\end{equation}
where $N$ is the dimension of the Hilbert space of our system.
In accordance with the required precision, if $C\leq 10^{-8}$, we identify the decomposition of $U_{\text{circ}}$ to be adequate for successful implementation of $U_{\text{goal}}$. We minimize the error $C$ with simple gradient descent without momentum, and brief experimentation suggested that different gradient descent variants performed very comparably on this task. It was numerically advantageous to update each parameter sequentially as opposed to updating all variables in a batch, using a dynamical representation of the left and right parts of the circuit (similar to the "sweeping" typical in MPS calculations). This was the variant of gradient descent that was used in most experiments.

\subsubsection{Approximation to the closest unitary through SVD}\label{sec:SVD}
Seeking an acceleration to simple gradient descent, we consider skipping several steps of the descent process by singular value decomposition (SVD). To motivate this decision, note that optimizing the parameters in a single $U_i$ is a unimodal optimization problem, and thus sufficiently many steps of gradient descent at one site will necessarily be equivalent to this SVD accelerator. The method is as follows. For any single-qubit unitary $U_i$ that we optimize, we trace out the subsystem of interest, $A_i$, for each iteration. Then, we find the closest unitary to $A_i$ through SVD \cite{Quek2022-td}. This is some $U_i$ such that the operator norm $\Vert A_i - U_i\Vert_2$ is minimized. The approximation goes as,
\begin{align}
    A_i = X \Sigma Y^{\dagger}\label{SVD} \\
    U_i = XY^{\dagger}
\end{align}
where \eqref{SVD} is the SVD operation. The new unitary $U_i$ is normalized so that we are still working with an $\text{SU}(2)$ matrix, and the Euler angles $\theta_1, \theta_2 \text{ and } \theta_3 $ are extracted and stored.

In a mathematically precise sense, we can say that the SVD does not help escape local minima; it only accelerates the same computation that gradient descent could do anyway with an appropriate learning rate and site schedule. This acceleration avoids the need to pick a learning rate, and we can thus view this as a more manifold-aware optimization method.

\subsubsection{Layer Calculation}
During implementation, we need to calculate the size (equivalently, depth) of the required circuit. This is a straightforward exercise in parameter counting that we repeat here for clarity.

In a system with $n$ qubits, the unitary $U$ is  $N \times N$, where $N=2^n$, and has $P = N^2 - 1 = 4^n - 1$ real parameters. According to our earlier decomposition, we segment the circuit into layers $S_iC_i$ of alternating single-qubit unitaries and CNOTs. A general single-qubit unitary carries three free parameters, so each layer appears to contribute $3n$ parameters from the local gates. But then the symmetries of the CNOT gates reduce the overall parameter count. Specifically, a CNOT is invariant under (i) conjugation by Z rotation on the control, and (ii) conjugation by X rotation on the target, i.e.,
\begin{align*}
R_Z(\theta)\otimes \mathbb{I} (\text{CNOT}) R_Z(\theta) \otimes \mathbb{I} =\\
\mathbb{I}\otimes R_X(\theta) (\text{CNOT}) \mathbb{I} \otimes R_X(\theta) = \text{CNOT}.
\end{align*}
These two symmetries effectively reduce 2 free parameters per CNOT.

\par One way of understanding this is that we could always choose to \textit{normalize} a circuit according to the following procedure:
\begin{enumerate}
    \item Writing each single qubit unitary in the first layer as $R_X(\theta_1)R_Y(\theta_2)R_Z(\theta_3)$ if it qubit will a control qubit in the next CNOT, or $R_X(\theta_1)R_Y(\theta_2)R_X(\theta_3)$ if it will be the qubit acted on by the CNOT. These are two different Euler angle decompositions.
    \item Commute the last rotation (either $R_Z$ or $R_X$) past the CNOT, altering the single-qubit unitaries in the second layer.
    \item Proceed in this way for all layers, until the last, which is not followed by a CNOT.
\end{enumerate}
Now all single-qubit unitaries have a canonical form with two real parameters, except for the last layer which has three real parameters, so that the total circuit with $\ell$ layers has $P = 2n(\ell - 1) + 3n = (2\ell + 1)n$ parameters. To express a generic unitary we then require $(2\ell + 1)n \ge N^2 - 1$, or
\begin{equation}\label{eq:ellCount}
\ell = \lceil{\frac{4^n - 1 - n}{2n}\rceil}.
\end{equation}

An alternate calculation would say that each CNOT \textit{loses} us two degrees of freedom; this symmetry can be calculated from the degeneracy of its spectrum. Since CNOT has a spectrum of $\{-1,1,1,1\}$, it has two degeneracies, which manifest as the loss of two effective parameters. It is worth considering the parameter loss due to different two-qubit coupling gates:
\begin{itemize}
    \item The CZ gate would have the same parameter loss of 2, as it is locally equivalent to CNOT.
    \item The Sycamore gate \cite{Harrigan2021} has a spectrum $\{1,-i,i,\exp(-i\pi/6)\}$ and no degeneracies; no effective parameters are lost. This is also true for a generic two-qubit gate.
    \item The SWAP gate has only two degeneracies in the spectrum, but it also fails to mix the left- and right-symmetric subspaces, leading to a total loss of six real parameters. This means \textit{that all} parameters from a layer of unitaries are lost, and, of course, $\text{SU}(2) + $ SWAP is not universal for computation.
\end{itemize}

In section \ref{sec:constraint} we discuss restricted circuit topologies, where we cannot make use of every qubit in every layer. Then the layer calculation is slightly altered.
\section{Results}
\begin{figure*}[ht]
\centering

\begin{subfigure}{0.47\textwidth}
\centering
\[
\begin{quantikz}
\lstick{\(q_1\)} & \gate{U_1} & \ctrl{1} & \gate{U_3} & \ctrl{1} & \gate{U_5} & \ctrl{1} & \gate{U_7} & \qw \\
\lstick{\(q_2\)} & \gate{U_2} & \targ{}  & \gate{U_4} & \targ{}  & \gate{U_6} & \targ{}  & \gate{U_8} & \qw
\end{quantikz}
\]
\caption{2-qubit unitary decomposition}
\label{fig:2qubitDecomp}
\end{subfigure}
\hfill
\begin{subfigure}{0.47\textwidth}
\centering
\[
\begin{quantikz}
\lstick{\(q_1\)} & \gate{U_1} & \ctrl{1}      & \gate{U_5}  & \ctrl{2} & \qw      & \gate{U_9}  & \ctrl{3} & \qw & \qw  \\
\lstick{\(q_2\)} & \gate{U_2} & \targ{}       & \gate{U_6}  & \qw      & \ctrl{2} & \gate{U_{10}} & \qw & \ctrl{1} & \qw  \\
\lstick{\(q_3\)} & \gate{U_3} & \ctrl{1} & \gate{U_7}  & \targ{}  & \qw      & \gate{U_{11}} & \qw & \targ{} & \qw  \\
\lstick{\(q_4\)} & \gate{U_4} & \targ{}  & \gate{U_8}  & \qw      & \targ{}  & \gate{U_{12}} & \targ{} & \qw & \qw 
\end{quantikz}
\]
\caption{One component motif of a 4-qubit unitary decomposition, to be repeated until $l=32$}
\label{fig:4qubitDecomp}
\end{subfigure}

\begin{subfigure}{1\textwidth}
\centering
\[
\begin{quantikz}[row sep=0.3cm]
\lstick{\(q_1\)} & \gate{U_1} & \ctrl{1} & \gate{U_7}  & \ctrl{3} & \qw & \qw & \gate{U_{13}} & \ctrl{5} &\qw &\qw &\gate{U_{19}} &\ctrl{2} &\qw &\qw &\gate{U_{25}} &\ctrl{4}  &\qw  &\qw\\
\lstick{\(q_2\)} & \gate{U_2} & \targ{}  & \gate{U_8}  & \qw      & \ctrl{3} & \qw & \gate{U_{14}} &\qw & \ctrl{2} &\qw &\gate{U_{20}}&\qw &\ctrl{4} &\qw &\gate{U_{26}}  &\qw &\ctrl{1}  &\qw\\
\lstick{\(q_3\)} & \gate{U_3} & \ctrl{1} & \gate{U_9}  & \qw      & \qw      & \ctrl{3} & \gate{U_{15}} &\qw &\qw & \ctrl{2} &\gate{U_{21}} &\targ{} &\qw &\qw &\gate{U_{27}}  &\qw &\targ{}  &\qw\\
\lstick{\(q_4\)} & \gate{U_4} & \targ{}  & \gate{U_{10}} & \targ{} & \qw & \qw & \gate{U_{16}}&\qw & \targ{} &\qw &\gate{U_{22}} &\qw &\qw &\ctrl{1} &\gate{U_{28}} &\qw  &\qw &\ctrl{2}\\
\lstick{\(q_5\)} & \gate{U_5} & \ctrl{1} & \gate{U_{11}} & \qw      & \targ{} & \qw & \gate{U_{17}} &\qw &\qw & \targ{} &\gate{U_{23}} &\qw &\qw &\targ{} &\gate{U_{29}}&\targ{}  &\qw  &\qw\\
\lstick{\(q_6\)} & \gate{U_6} & \targ{}  & \gate{U_{12}} & \qw      & \qw & \targ{} & \gate{U_{18}} & \targ{} &\qw &\qw &\gate{U_{24}} &\qw &\targ{} &\qw &\gate{U_{30}}  &\qw  &\qw &\targ{}
\end{quantikz}
\]
\caption{The repeated motif for a 6-qubit unitary decomposition}
\label{fig:6qubitDecomp}
\end{subfigure}

\caption{Circuit decompositions for 2-, 4-, and 6-qubit unitaries.}
\label{fig:allDecomps}
\end{figure*}
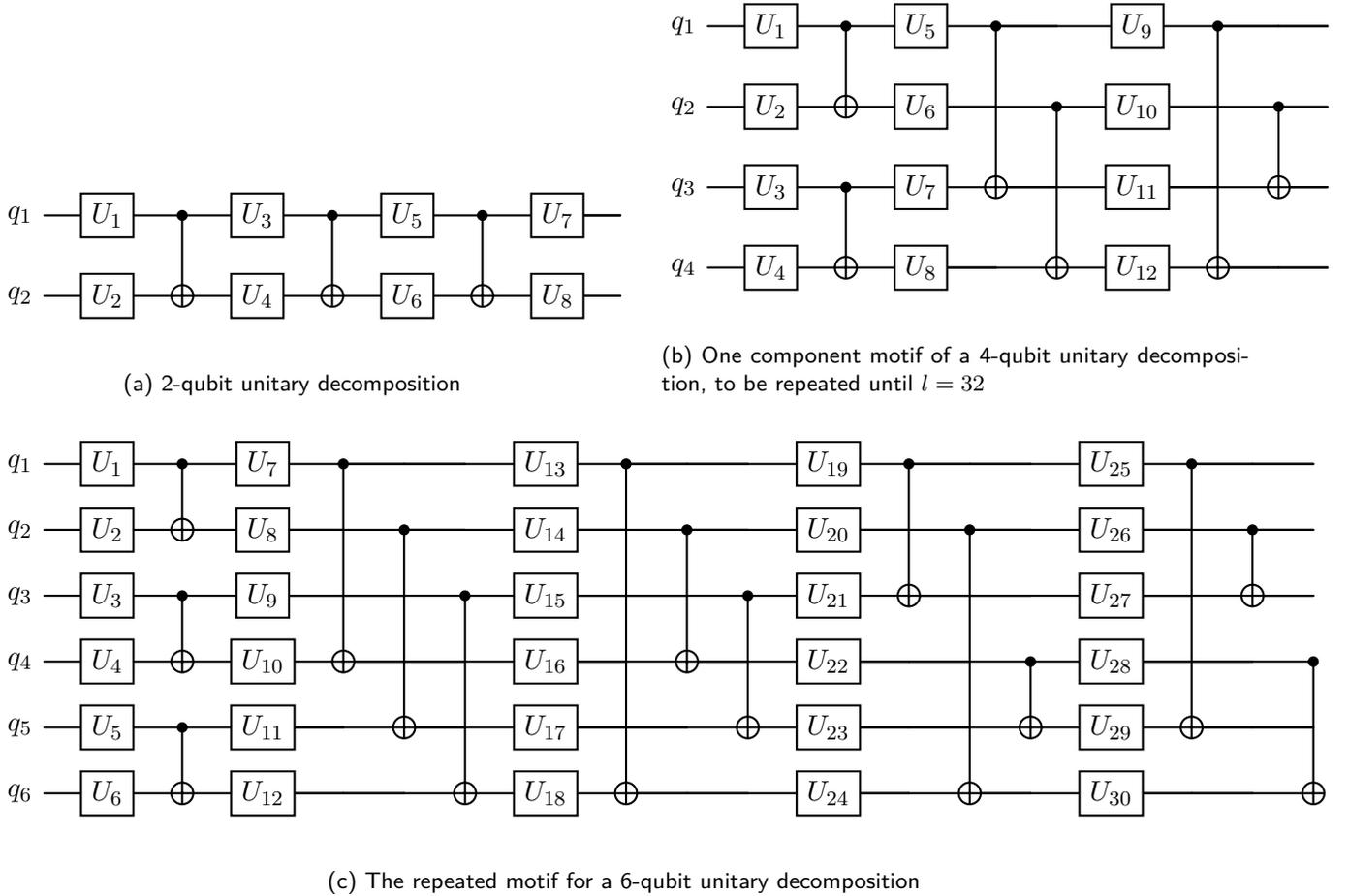

The primary question we want to investigate and address is the low success rate of a random combinatorial search in the case of adequate parameterization using the gradient descent method in optimization and, secondarily, its efficiency. The "cleanest" topology to study first was an even number of qubits in fully connected hardware, where we can exactly fill each layer of CNOTs with arbitrary connectivity. Therefore, we investigate our method for such two-qubit, four-qubit and six-qubit systems. In this section, we first describe the circuit decompositions for all system sizes. In particular, we elaborate on the importance and reasoning behind the choice of circuit topologies in bigger systems. Following that, we present the primary results we obtained by running the numerics across these different system sizes. Finally, we conclude this section with systems that do not have full connectivity and present conclusions on the effect of these constraints on the results.
\subsection{Circuit decomposition and choice of topology}
We demonstrate the choice of circuit design with two, four and six-qubit systems.

The two-qubit unitary needs three layers for adequate parameterization according to \ref{eq:ellCount}, and for that, there is a unique topology (see figure \ref{fig:2qubitDecomp}). This is of course well-studied, and indeed an optimal and exact decomposition of a two-qubit unitary is known, see e.g. \cite{Vatan2004-ik,Vidal2004-np}.

Moving on to the four-qubit unitary, which needs 32 layers according to \ref{eq:ellCount}, we now have a nontrivial question of picking the topology. Our first na\"ive strategy was to simply repeat the motif in \ref{fig:4qubitDecomp}.



Why this motif? Up to permutations of qubits, it is the \textit{unique} three-layer motif that directly connects all the qubits in the system to each other. 
\begin{enumerate}
    \item (1,2), (3,4)
    \item (1,3), (2,4)
    \item (1,4), (2,3)
\end{enumerate}
Believing that good mixing would be necessary for good trainability of the circuit, we wanted to avoid repeating a connection, so after these three pairings have been done, we go back to the least-recently used -- and repeat the cycle of three. This decision turned out to be fortuitous and likely optimal!

For six-qubit unitaries, we need $\ell =341$, according to \ref{eq:ellCount}. The choice of topology for this case is quite non-trivial. Here, we have ${}^{6}C_2 = 15 $ ways of picking out the pair of qubits for the CNOTs to act on in each layer. The order of these CNOTs becomes important in the decomposition, as certain patterns in the choice of subsystems can end up effectively reducing the number of parameters. 
    

Let us elaborate on such a scenario with an example. Consider that the same two qubits are paired together for four layers consecutively, that is, three rounds of CNOT pairings. Here, we end up using $3\cdot 8 - 2\cdot 3 = 18$ parameters to build what is effectively a unitary that acts only on two qubits. However, a general two-qubit unitary requires just 15 parameters. The surplus parameters introduce redundancies in the circuit, leading to symmetries, i.e., directions in parameter space along which the output unitary remains unchanged. These symmetries reduce the number of effective degrees of freedom available to the rest of the circuit. As a result, the system will not have enough parameters and will not converge, despite appearing to have an adequate number of layers. This ultimately means that there are certain CNOT configurations for which the circuit will fail to converge. 

In \cite{Ashhab2024-yw}, this probability of failure, or rather the percentage of configurations that converge, was found through brute-force, random combinatorial search. The failed configurations are likely cases of effective underparameterization as described above. We give further numerical justification of this belief in Appendix \ref{app:failure-prob}.

We made a more detailed investigation on the underparameterized regime of circuits for the bigger systems — four-qubit and six-qubit systems (see Appendix \ref{app:under}). This was considered in an effort to reduce the CNOT count further. On examining the effect of this condition on convergence, we observe that the cost function plateaus at a much higher error than usable, even when the number of layers is only slightly short of the required number. As the benefit in the reduction of gates does not compensate well enough for the loss in precision, we conclude that underparameterization is not practically useful in this setting


We continue to follow the logic that to ensure convergence, we require rapid mixing, with no reduction in the number of parameters. One way to do this is to make sure that each combination of CNOT pairing is only repeated in the circuit skeleton as far away as possible. Now, we must find the maximum number of other such combinations that can be inserted before repetition becomes inevitable. This problem is a well-known graph theory problem, referred to as the 1-factorization of a k-node graph. The maximum number of combinations corresponds to the number of 1-factors, which is given by $k-1$. In the case of six nodes, this gives five such combinations. These five combinations are listed below. 

\begin{enumerate}
    \item (1,2), (3,4), (5,6)
    \item (1,4), (2,5), (3,6)
    \item (1,6), (2,4), (3,5)
    \item (1,3), (2,6), (4,5)
    \item (1,5), (2,3), (4,6)
\end{enumerate}
It is important to note that this set of combinations is not unique, especially when the number of nodes is large. As before, after these five pairings, the CNOTs are repeated throughout the circuit in this order. This circuit is shown in figure \ref{fig:6qubitDecomp}.
\begin{figure*}[t]
    \centering 
    \begin{subfigure}{0.48\textwidth}
        \centering
        \includegraphics[width=\linewidth]{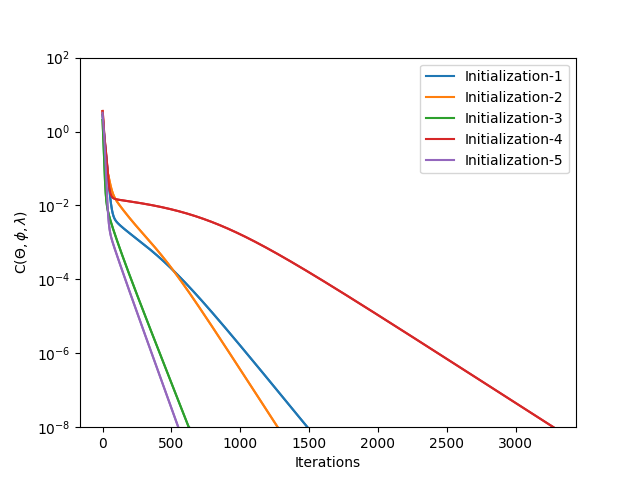}
        \caption{Convergences of the same $4\times4$ target unitary initialized with five different parameter sets. The y-axis is in log scale.}
        \label{fig:2 qubit different unitaries}
    \end{subfigure}
    \hfill
    \begin{subfigure}{0.48\textwidth}
        \centering
        \includegraphics[width=\linewidth]{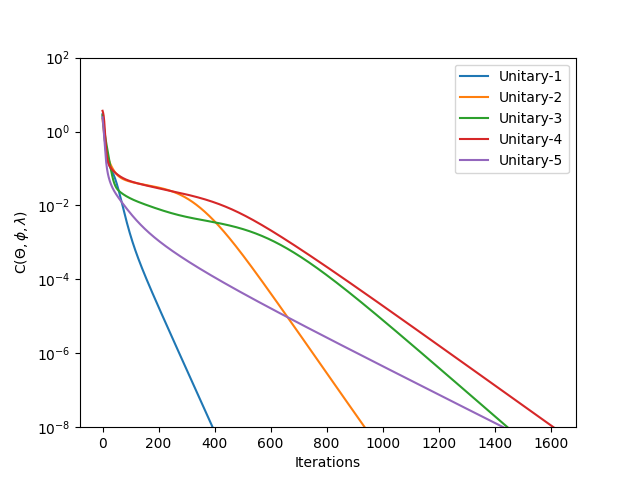}
        \caption{Convergences of five random $4\times4$ target unitaries optimized using the gradient-based algorithm. The y-axis is in log scale.}
        \label{fig:2 qubit different initializations}
    \end{subfigure}

    \caption{Comparison of convergence behavior for different initializations (left) and different target unitaries (right).}
    \label{fig:2qubit-side-by-side}
\end{figure*}
\subsection{Numerics}
After considering the choice of circuit topologies, the optimization algorithm was run for around 200 two-qubit, 100 four-qubit and 30 six-qubit unitaries. Here we note that the results stated in this paper held without fail for all of these trials, corroborating the numerical evidence in \cite{kiani2020learningunitariesgradientdescent}. The target unitaries for all system sizes were generated Haar-randomly using a function from the \texttt{scipy} library. During the runs, we optimized the parameters of the single-qubit unitaries using gradient descent for the following cases: (a) arbitrary target unitaries, (b) the same random target unitary with the decomposition initialized with different sets of parameters.

We observed that the circuit reliably converges in both cases, avoiding any local minima. Although certain initializations lead to faster convergence than others, this does not affect the overall consistency in convergence. Hence, we proceeded with random initialization for all subsequent runs. These findings are illustrated in figure \ref{fig:2 qubit different unitaries} and figure \ref{fig:2 qubit different initializations}. Even as the system size increases, i.e. for four-qubit and six-qubit systems, these results continued to hold (see appendix \ref{app:numeric-data}).
\subsection{Handling Connectivity Constraints}\label{sec:constraint}
So far, all the circuit topologies we have considered assume full qubit connectivity. However, practical quantum hardware typically imposes connectivity constraints that limit the combination of qubits that can be directly entangled. To evaluate the performance of our method under such conditions, we now examine circuits with restricted connectivity.  
Despite the connectivity constraints, both cases show reliable convergence with sufficient parameterization, while underparameterized circuits cause the cost function to plateau (see appendix \ref{app:low-connect}).

This means that we have obtained a parameter-optimal method for generic unitary synthesis in the case of circuits with low connectivity. This is especially advantageous over alternative methods of unitary synthesis, such as general-purpose quantum compilers or exact synthesis techniques, which are not parameter-optimal. Further, one can expect the quantum compilers to give a circuit with a higher number of parameters and CNOT count due to the suboptimal decomposition methods they use. Another advantage of this method is that it works even when the actions of the target unitaries are not specified for all states. One can still measure the cost function for the subset of specified actions and perform gradient descent-based optimization for them.

\section{Conclusion}
Despite numerical evidence that optimal solutions can be reached with adequate parameterization, random combinatorial search exhibits a very low success probability, particularly for larger systems. We attribute this observation to the increasing importance of the CNOT configuration with system size: symmetries between circuit layers can effectively reduce the number of independent parameters, obstructing convergence. We address this by prescribing circuit skeletons that avoid such parameter deficiencies and enable reliable convergence under gradient-descent optimization. Further, the gradient descent-based optimization method remains robust even under limited connectivity, making it particularly well-suited to be used on realistic quantum hardware.

This method also offers a practical and efficient framework for unitary synthesis in contrast to alternatives such as exact synthesis algorithms, which assume full connectivity, or standard quantum compilers, which may not yield parameter-efficient circuits.

\section{Acknowledgements}

We thank David Gosset for his support and insights. JG also sincerely thanks Keshav Das Agarwal for many fruitful discussions.\\
Research at Perimeter Institute is supported in part by the Government of Canada through
the Department of Innovation, Science and Industry Canada and by the Province of Ontario
through the Ministry of Colleges and Universities. Both authors were supported by Discovery Grant No. RGPIN-2019-0419.

\bibliographystyle{plain}
\bibliography{refs}

\onecolumn\newpage
\appendix

\section{Reliable convergence of 4-qubit and 6-qubit systems}\label{app:numeric-data}
As in the case of two qubit systems, we analyze the convergence of (a) random target unitaries, (b) the same target unitary initialized with different sets of parameters, for four qubit and six qubit systems. In (a), we pick five random target unitaries and run the algorithms for all of them till the cost function converges, indicating that the circuit now represents the target. For (b), we choose five different initializations for the same target unitary and compare the convergences.
\subsection{4-qubit system}
\begin{figure}[H]
    \centering
    \begin{minipage}{0.45\textwidth}
        \centering
        \includegraphics[width=\linewidth]{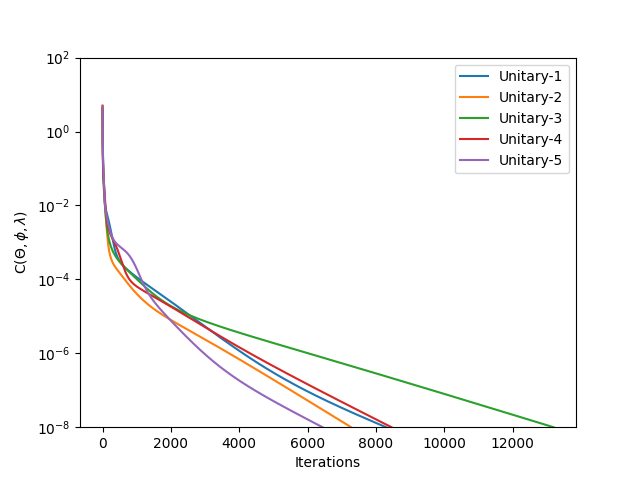}
        \caption{Convergences of five random $16 \times 16 $ target unitaries optimized through the gradient-based algorithm for $\ell=32$. The y-axis is in log scale.}
        \label{fig:qubit4diffUnitary}
    \end{minipage}
    \hfill
    \begin{minipage}{0.45\textwidth}
        \centering
        \includegraphics[width=\linewidth]{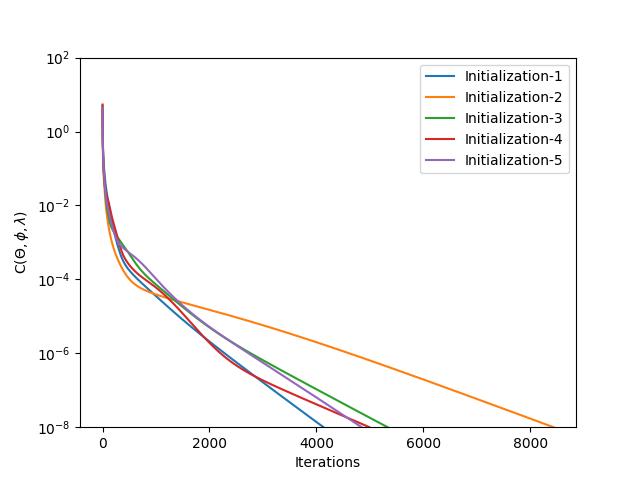}
        \caption{Convergences of the same $16 \times 16 $ target unitary initialized with five different sets of parameters for $\ell=32$. The y-axis is in log scale.}
        \label{fig:qubit4diffInit}
    \end{minipage}
\end{figure}

    

\subsection{6-qubit system}
\begin{figure}[H]
    \centering
    \begin{minipage}{0.45\textwidth}
        \centering
        \includegraphics[width=\linewidth]{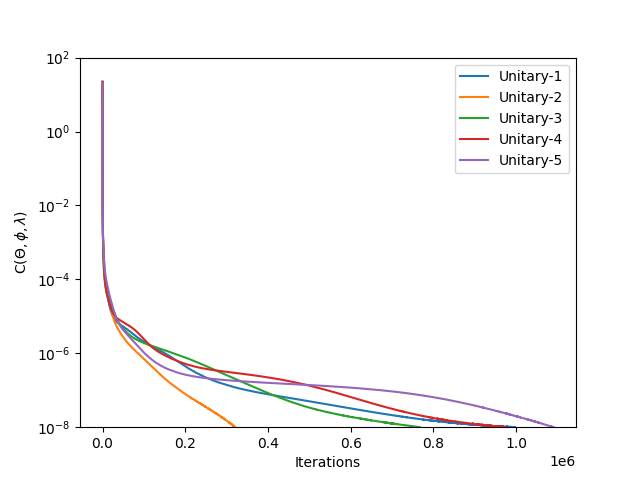}
        \caption{Convergences of five random $64 \times 64 $ target unitaries optimized through the gradient-based algorithm for $\ell=341$.The y-axis is in log scale.}
        \label{fig:qubit6diffUnitary}
    \end{minipage}
    \hfill
    \begin{minipage}{0.45\textwidth}
        \centering
        \includegraphics[width=\linewidth]{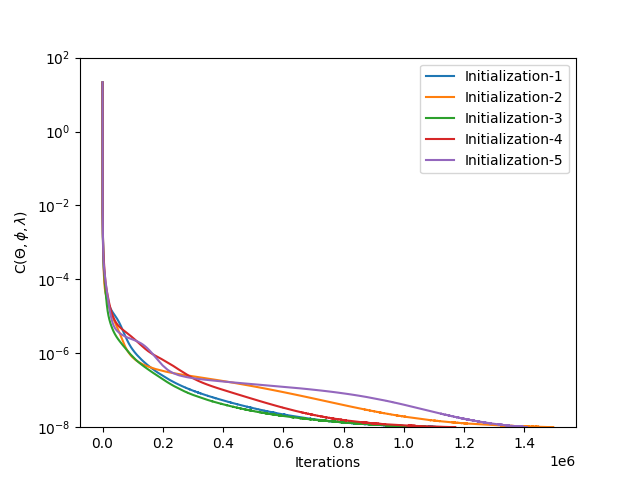}
        \caption{Convergences of the same $64 \times 64 $ target unitary initialized with five different sets of parameters for $\ell=341$.The y-axis is in log scale.}
        \label{fig:qubit6diffInit}
    \end{minipage}
\end{figure}
Both analyses give the following conclusion: The method robustly finds the circuit for any generic unitary regardless of initial parameters.
\section{Underparameterized Circuits}\label{app:under}
After ensuring that we can find a circuit with sufficient accuracy in the case of adequate number of layers, the next objective becomes minimizing the number of CNOTs in the constructed circuit. For this, we study the underparameterized regime for bigger systems, examining its effect on convergence. The goal is to determine whether convergence can still be achieved with fewer parameters, potentially reducing the CNOT count and improving practical implementability. To this end, we explore the underparameterized circuits for the four-qubit and six qubit system. It is important to note that for six qubit systems, the definition of an underparameterized circuit is refined by the choice of circuit topology as well. We observe that the cost function plateaus at $\approx 10^{-3}$ in the best case for underparameterized circuits, which is well before reaching the desired precision of $10^{-8}$, even when the number of layers is only slightly short of the required number. As the benefit in the reduction of gates does not compensate well enough for the loss in precision, we conclude that underparameterization is not very useful. Additionally, we look at circuits that are slightly over-parameterized. Interestingly, the over-parameterized circuits converge faster than the adequately parameterized circuits, suggesting that a small parameter surplus aids in convergence. 

\begin{figure}[H]
    \centering
    \begin{minipage}{0.45\textwidth}
        \centering
        \includegraphics[width=\linewidth]{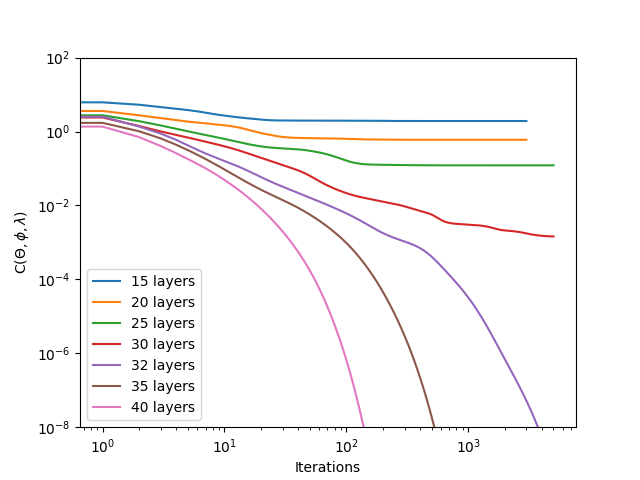}
    \caption{Convergences of the same 4 qubit ($16 \times 16$) target unitary for different $\ell$ values. Both x-axis and y-axis are in log scale to aid in comparing various $\ell$ values.}
    \label{fig:qubit4diffLayers}
    \end{minipage}
    \hfill
    \begin{minipage}{0.45\textwidth}
        \centering
        \includegraphics[width=\linewidth]{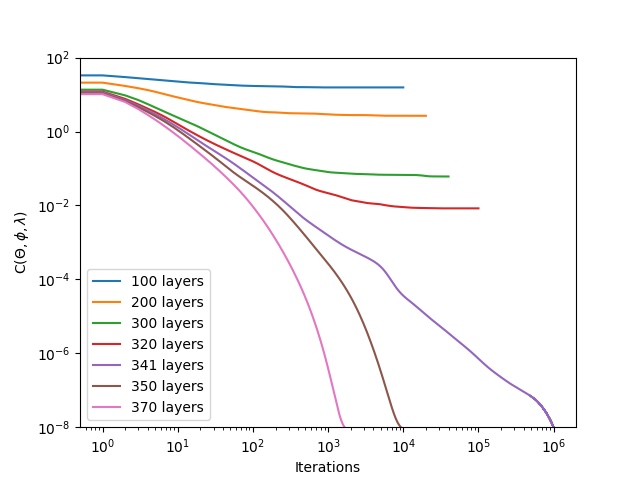}
    \caption{Convergences of the same 6 qubit ($64 \times 64$) target unitary for different $\ell$ values. Both x-axis and y-axis are in log scale to aid in comparing various $\ell$ values.}
    \label{fig:qubit6diffLayers}
    \end{minipage}
\end{figure}

Thus, our final conclusion is that the success of optimization is guaranteed under adequate parameterization, and the underparameterized circuits fail to converge(see figures \ref{fig:qubit4diffLayers} and \ref{fig:qubit6diffLayers}). 

\subsection{Other optimization techniques}
Given that simple gradient descent (or our accelerated variant, see Section \ref{sec:SVD}) failed to find circuits of useful precision at even very mild underparameterization, a natural question was whether more advanced search strategies could succeed. While there are many approaches and algorithms to investigate for this hard problem, our experiments focused on whether a straightforward {\em modification or extension} of gradient descent could suffice. The most obvious way to do this seemed to allow modification of the CNOT topology during search. We conducted brief investigations into several related variants.
\begin{itemize}
    \item Periodically during optimization, a CNOT layer is exchanged for a different arrangement. If the reconstruction accuracy improves, keep it.
    \item Alternately: keep it with some other probability, even if the accuracy worsens. This is a form of simulated annealing, and the acceptance probability should reduce towards zero in order to ensure convergence.
    \item Or as another possibility, instead of allowing an arbitrary CNOT change, only allow changes that don't create underparameterized ``redundant" configurations.
    \item At a given step, instead of trying just one alternate CNOT configuration, try every change at every level, to see if there is any similar CNOT configuration that improves.
    \item Instead of just changing the CNOT layer, change the CNOT, then re-fit across several steps of gradient descent. If this is better than the original circuit, then accept. This would allow for changes that overall improve but require parameter adjustments to work.
    \item Instead of just changing one CNOT layer, scramble the CNOTs and all single-qubit unitaries on a pair of adjacent layers, and re-fit them all; see if this produces an improved circuit. If it does, accept.
\end{itemize}
None of these seemed to provide a noticeable improvement over the simple gradient descent searches. It seems very difficult to make discrete jumps in topology cooperate with the continuous nature of gradient descent, and so we believe that approaches like continuous relaxations of circuit topologies are likely more promising in this regard.

\section{Circuits with low connectivity}\label{app:low-connect}
\begin{figure}[H]
    \centering
    \includegraphics[width=0.35\linewidth]{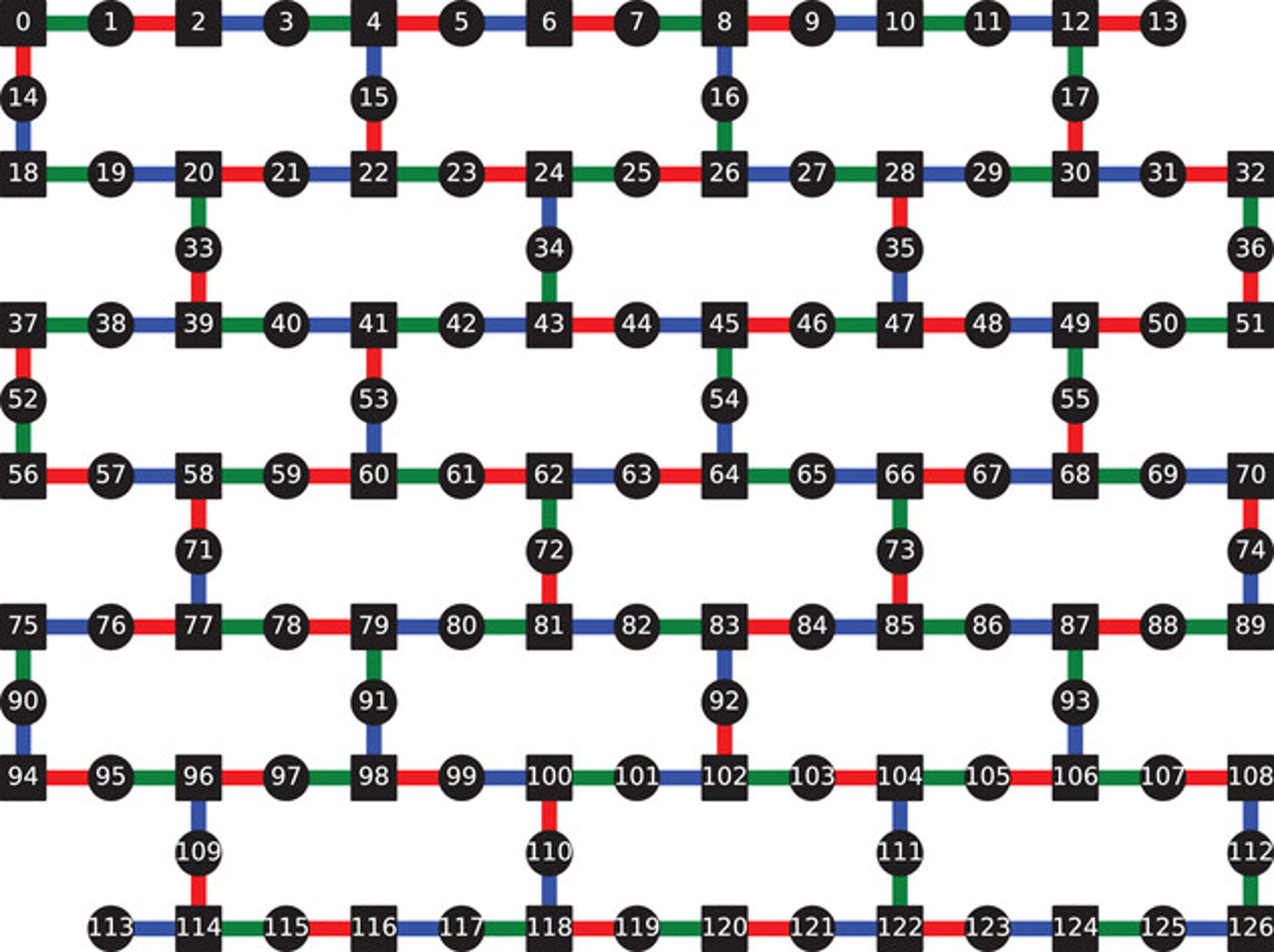}
    \caption{127 qubit IBM Chip, Adv Quantum Tech, Volume: 8, Issue: 1, First published: 09 October 2024, DOI: (10.1002/qute.202400239) 
}
    \label{fig:IBM}
\end{figure}

We investigate two representative cases of low connectivity, which are sub-systems of the 127-qubit IBM chip(see figure \ref{fig:IBM}).The immediate significance of low connectivity is that not all qubits are connected to each other, which means there is a reduction in the total number of parameters in the circuit in a given layer. Nevertheless, as long as the circuit remains adequately parameterized (by increasing the number of layers), there is no fundamental reason to expect a failure of the optimization. Note that since the total number of parameters remains the same, the increase in the number of layers has no effect on the required number of gates.
\subsection{Case 1: Qubits 3,4,5,15}
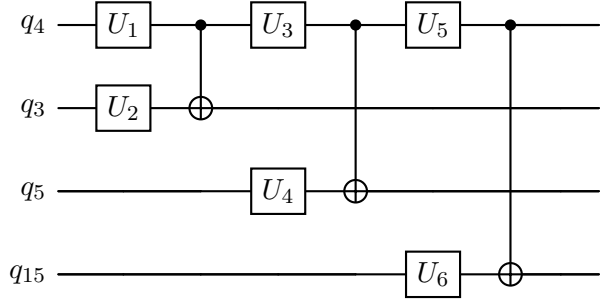
\begin{figure}[H]
    \centering  
\[
\begin{quantikz}
\lstick{\(q_4\)} & \gate{U_1} & \ctrl{1} & \gate{U_3}  &\ctrl{2} &\gate{U_5} &\ctrl{3} &\qw &\qw \\
\lstick{\(q_3\)} & \gate{U_2} & \targ{} &\qw &\qw &\qw &\qw &\qw &\qw\\
\lstick{\(q_5\)} & \qw & \qw      & \gate{U_4}  & \targ{}  &\qw &\qw &\qw &\qw\\
\lstick{\(q_{15}\)} & \qw & \qw     & \qw      & \qw &\gate{U_6} &\targ{} &\qw &\qw
\end{quantikz}
\]
\caption{Circuit decomposition for a $16 \times 16$ unitary that respects connectivity constraints of qubits 3,4,5,15.}
    \label{IBM 34515}
\end{figure}

We consider qubits 3, 4, 5, and 15 of the chip with the following connectivity constraint: qubit 4 can only be coupled to one of 3, 5, or 15 at the same time; and 3, 5, and 15 are not connected to each other. Effectively, we now have two single-qubit unitaries and one CNOT in each layer. Consequently, the minimum number of layers required for convergence becomes 64. The circuit structure is illustrated in figure \ref{IBM 34515}.

\begin{figure}[h]
    \centering
    \begin{minipage}{0.45\textwidth}
        \centering
        \includegraphics[width=\linewidth]{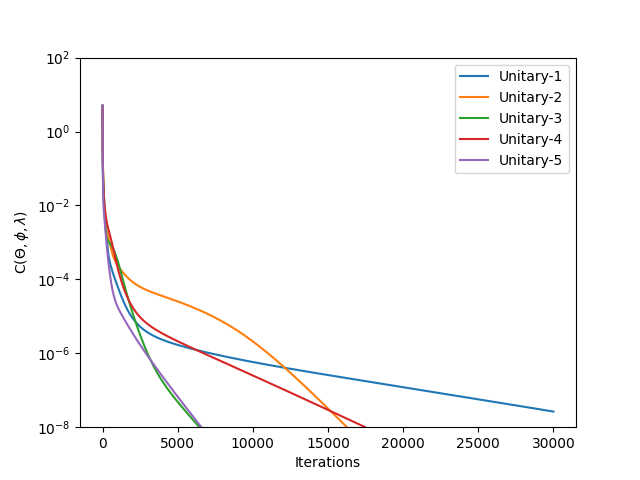}
        \caption{Convergences of five random 16 $\times$ 16 target unitaries for $\ell$ = 64 with the y-axis in log scale. The decomposition respects the hardware constraints imposed for qubits 3,4,5,15.}
        \label{fig:ibmDiffUnitary}
    \end{minipage}
    \hfill
    \begin{minipage}{0.45\textwidth}
        \centering
        \includegraphics[width=\linewidth]{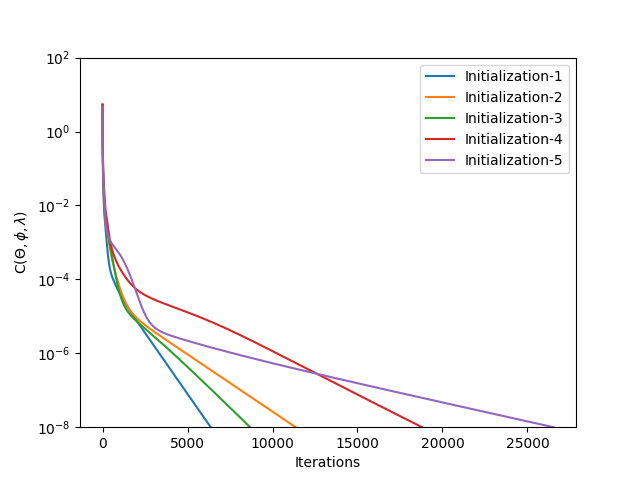}
        \caption{Convergences of the same 16 $\times$ 16 target unitary initialized with five different sets of parameters for $\ell$ = 64. The y-axis is in log scale. The decomposition respects the hardware constraints imposed for qubits 3,4,5,15.}
        \label{fig:ibmDiffInit}
    \end{minipage}
\end{figure}

\begin{figure}[H]
    \centering
    \includegraphics[width=0.45\linewidth]{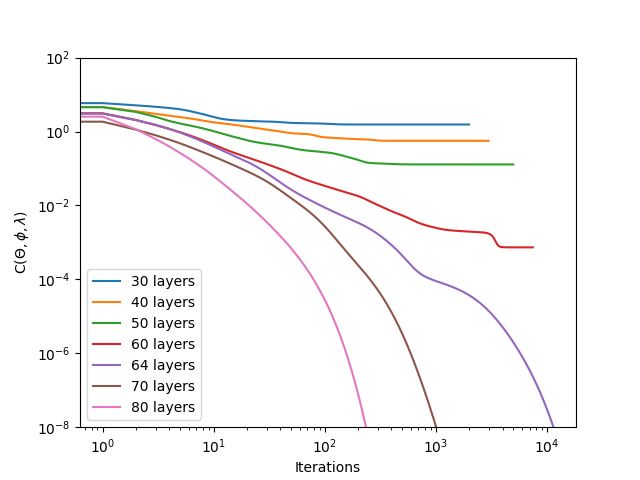}
    \caption{Convergences of the same $16 \times 16$ target unitary for different $\ell$ values. The decomposition respects the hardware constraints imposed for qubits 3,4,5,15. Both x-axis and y-axis are in log scale to aid in comparing various $\ell$ values.}
    \label{fig:ibmDiffLayers}
\end{figure}

As before, the graphs \ref{fig:ibmDiffUnitary} and \ref{fig:ibmDiffInit} discuss the reliable convergence in the case of random target unitaries and the same unitary with different initializations respectively. Meanwhile, graph \ref{fig:ibmDiffLayers} discusses the saturation of the cost function in the case of underparameterization and presents a few cases of adequate parameterization for comparison. Convergence is reliable with sufficient parameterization, whereas underparameterized circuits cause the cost function to plateau. 

\subsection{Case 2: Qubits 0,1,2,3}
\begin{figure}[H]
    \centering
\[
\begin{quantikz}[row sep=0.5cm, column sep=0.7cm]
\lstick{\(q_0\)} & \gate{U_1} & \ctrl{1} &\qw &\qw &\qw \\
\lstick{\(q_1\)} & \gate{U_2} & \targ{} &\gate{U_5} &\ctrl{1} &\qw  \\
\lstick{\(q_2\)} & \gate{U_3} &\ctrl{1}      & \gate{U_6}  & \targ{} &\qw \\
\lstick{\(q_3\)} &\gate{U_4}&\targ{} & \qw & \qw  &\qw 
\end{quantikz}
\]
\caption{Circuit decomposition for a $16 \times 16$ unitary that respects connectivity constraints of qubits 0,1,2,3.}
    \label{IBM 0123}
\end{figure}
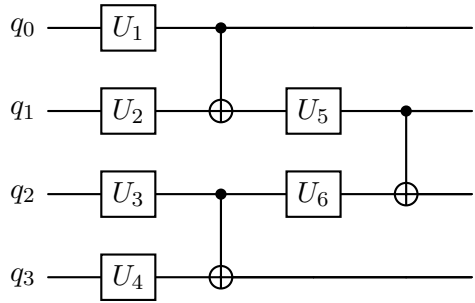
 Next, we consider qubits 0,1,2,3 with the following connectivity constraint: only adjacent qubits can be connected, and each can be connected to only one other qubit at a time. The circuit is as shown in figure \ref{IBM 0123} with $\ell= 42$ and the results are illustrated in the three graphs below.
 \begin{figure}[H]
    \centering
    \includegraphics[width=0.45\linewidth]{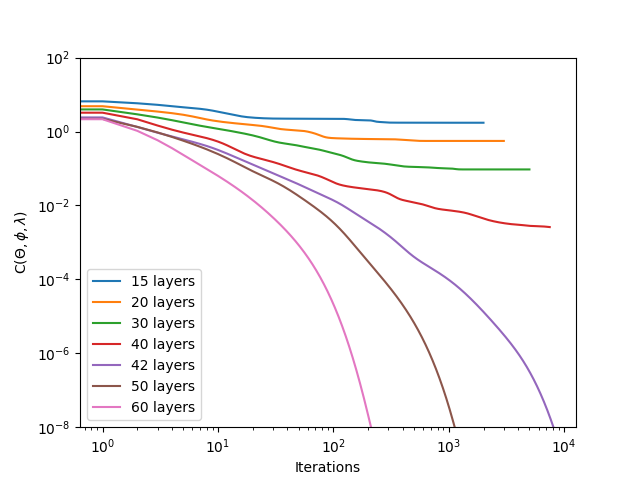}
    \caption{Convergences of the same $16 \times 16$ target unitary for different $\ell$ values. The decomposition respects the hardware constraints imposed for qubits 0,1,2,3. Both x-axis and y-axis are in log scale to aid in comparing various $\ell$ values.}
    \label{fig:IBM2diffLayers}
\end{figure}
\begin{figure}[h]
    \centering
    \begin{minipage}{0.45\textwidth}
        \centering
        \includegraphics[width=\linewidth]{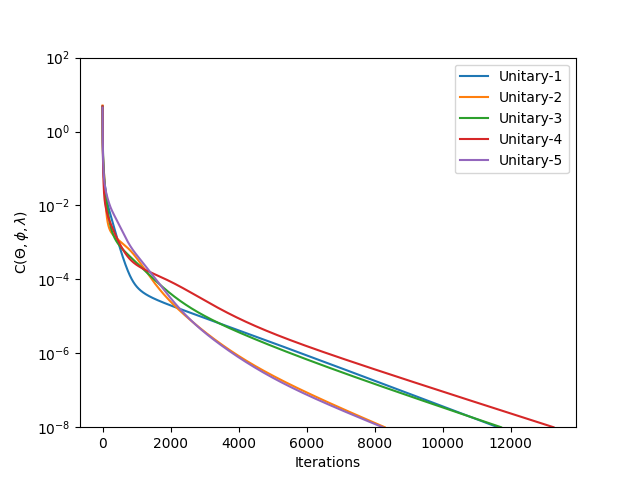}
    \caption{Convergences of five random 16 $\times$ 16 target unitaries for $\ell$ = 42 with the y-axis in log scale. The decomposition respects the hardware constraints imposed for qubits 0,1,2,3.}
    \label{fig:IBM2diffUnitary}
    \end{minipage}
    \hfill
    \begin{minipage}{0.45\textwidth}
        \centering
        \includegraphics[width=\linewidth]{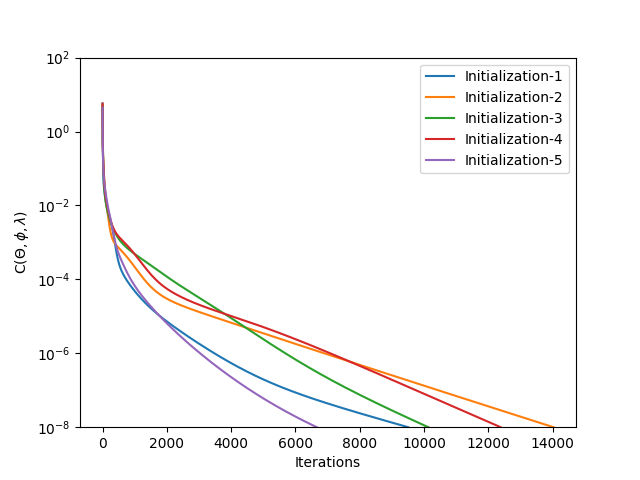}
    \caption{Convergences of the same 16 $\times$ 16 target unitary initialized with five different sets of parameters for $\ell$ = 42. The y-axis is in log scale. The decomposition respects the hardware constraints imposed for qubits 0,1,2,3.}
    \label{fig:IBM2diffInit}
    \end{minipage}
\end{figure}

Thus, we can conclude that even in the case of a decomposition with low connectivity, we obtain reliable convergence, regardless of initial parameters, as long as there is adequate parameterization—the cost function plateaus in the case of underparameterization.

\section{Calculation of Failure Probability in Random Circuits}\label{app:failure-prob}
We hypothesize that a poor pairing structure can explain all the failures observed in \cite{Ashhab2024-yw}. To corroborate this, we can see whether theory correctly replicates the frequency of failures. For a given circuit topology, we can compute its effective parameters using the principles outlined above. Then if this effective parameter count is less than $4^n-1$, failure is (generically) necessary.

Any block of three CNOTs coupling the same pair of qubits in a row creates an arbitrary two-qubit gate, giving only 15 parameters instead of the $6 + 4\cdot 3 = 18$ we would hope for, losing us three effective parameters. Any subsequent gates will lose another 4. Similarly, a block of sufficiently many CNOTs that keeps coupling the same set of three qubits will max out at $4^3 - 1 = 63$. For CNOTs on an $n=3$ circuit, we can reasonably work this out in closed form.

With $N=14$ CNOT gates (using the notation of \cite{Ashhab2024-yw} - see their Figure 2), any triple repeat will lead to underparameterization, so we count length-14 words on $ABC$ with no triple letter. This gives a success rate of $\frac{1526976}{3^{14}} \approx 31.9\%$. When $N=15$, we can have at most one triple repeats, but no quadruple string, giving $\frac{10040832}{3^{15}} \approx 69.98\%$. And repeating the calculation for $N=16$, we can have a triple repeat or quadruple repeat, but no two, giving $\frac{37327104}{3^{16}} \approx 86.7\%$ success rate. This agrees very well with the numbers reported in \cite{Ashhab2024-yw}.

For larger $n$, the case work on different ways to assign qubits and expected numbers of qubits in different blocks becomes unwieldy to do on paper, but is easily calculated by a computer program. We find excellent agreement with the theory; for $n=4$, a < 1\% success chance with $N=61$ CNOT gates, a \~ 50\% success rate for $N=64$, and high success rates for $N \ge 67$. The $n=5$ agreement is similar.

Asymptotically, we can encounter a loss of parameters as soon as any CNOT is placed on the same pair of qubits three times in a row. If the circuit topology is selected by random pairings at each layer, the probability of repeating one pair three times in a row in a given triple of layers is $O(1/N)$. (A $1/N$ chance that the second layer pairs qubit 1 to the same qubit as the first layer, and another factor $1/N$ for the third layer; but there are $N$ qubits, so that the total probability is $\sim 1/N$.) Given that the necessary circuit depth goes as $O(2^N / N)$, while the per-layer failure rate is $O(1/N)$, the probability of no failures (being adequately parameterized at the optimal depth) will necessarily drop doubly exponentially, $2^{-2^{O(N)}}$.
\end{document}